\documentclass[conference]{IEEEtran}

\usepackage{graphicx} 
\usepackage{subcaption}
\usepackage{amsmath}
\usepackage{url}
\usepackage{wrapfig}
\graphicspath{{images}}

\begin{document}
\title{Enforceable Data Sharing Agreements\\Using Smart Contracts}

\author{
  \IEEEauthorblockN{
    Harsh Desai\IEEEauthorrefmark{1},
    Kevin Liu\IEEEauthorrefmark{2},
    Murat Kantarcioglu\IEEEauthorrefmark{1},
    Lalana Kagal\IEEEauthorrefmark{2}
  }
  \\
  \IEEEauthorblockA{
    \IEEEauthorrefmark{1}School of Computer Science
    \\
    The University of Texas at Dallas
    \\
    \texttt{\{hbd140030, muratk\}@utdallas.com}}
    \\
    \IEEEauthorblockA{\IEEEauthorrefmark{2}Computer Science and Artificial Intelligence Lab
    \\
    Massachusetts Institute of Technology
    \\
    \texttt{kyliu@mit.edu}
    \\
    \texttt{lkagal@csail.mit.edu}
  }
}

\maketitle
\begin{abstract} \label{abstract}
As more and more data is collected for various reasons, the sharing of such data becomes paramount to increasing its value. Many applications ranging from smart cities to personalized health care require individuals and organizations to share data at an unprecedented scale. Data sharing is crucial in today's world, but due to privacy reasons, security concerns and regulation issues, the conditions under which the sharing occurs needs to be carefully specified. Currently, this process is done by lawyers and requires the costly signing of legal agreements. In many cases, these data sharing agreements are hard to track, manage or enforce. In this work, we propose a novel alternative for tracking, managing and especially enforcing such data sharing agreements using smart contracts and blockchain technology. We design a framework that generates smart contracts from parameters based on legal data sharing agreements. The terms in these agreements are automatically enforced by the system. Monetary punishment can be employed using secure voting by external auditors to hold the violators accountable. Our experimental evaluation shows that our proposed framework is efficient and low-cost.
\end{abstract}

\section{Introduction} \label{intro}
In the era of big data, the amount of data existing in the world is rapidly increasing. Furthermore, this data needs to be shared for various critical tasks. For example, cities can share the data they collected about transport patterns with different stake holders such as transportation companies to reduce the traffic jams during rush hour \cite{traffic-jams}. First responders and aid workers share data between them to better coordinate disaster relief efforts \cite{disaster-relief}. Of course, the importance of data sharing can be seen in many other scenarios such as healthcare where patient data can be shared among institutions for improving treatment decisions.

In many use cases, a basic data sharing scenario would involve a data requester (e.g., taxi company), a data provider (e.g., the city) and the data itself. Data sharing can be a constructive concept to improve collaboration, analyze and utilize data to generate significant value, and maintain accountability and transparency. At the same time, the misuse of such shared data could create significant challenges. To prevent misuse, reduce liability, and comply with regulations, organizations may wish to sign legal agreements before sharing data. For example, the data distributer would likely wish to be notified if the data requester's data server is breached, and these conditions could be specified via a data sharing agreement. Currently, enforcing such agreements may require costly litigation, and managing a large number of such legal data sharing agreements is challenging. Therefore, new techniques are required to move the paper based legal agreements to digital world where such agreements could be enforced automatically and efficiently. To address this challenge, we propose using smart contracts built on the blockchain that can achieve automatic enforcement via monetary payments.

Blockchain, popularly introduced as part of Bitcoin in 2008, is based on the concept of distributed ledger technology for transactions to be stored globally as blocks \cite{bitcoin}. Since its initial perception as a cryptocurrency, blockchain technology has evolved to more complicated systems, which can now transfer data as well. Ethereum is the predominant blockchain technology used in data transfers, and introduced the concept of smart contracts or crypto-contracts. These contracts are snippets of digital code that control and permission the transfer of cryptocurrencies or data over the blockchain \cite{ethereum}. Similar to the way a traditional legal contract defines rules and penalties in an agreement between two parties, a smart contract defines similar rules with an added benefit of being able to enforce these obligations. Since contracts are on a distributed public ledger and all transactions are known to all, the requirement of a trusted third party is removed. This also gives an advantage to the data sharing scenario of maintaining a permanent audit trail. The immutability property of distributed decentralized ledgers provides transparency and protection of vital data.

In this work, we propose a framework using Ethereum smart contracts where data can be shared in exchange for payment, and penalties can be enforced for breaching data sharing agreements. Since Ethereum smart contracts run on the Ethereum Virtual Machine (EVM), which is Turing complete, a smart contract can spawn another smart contract with the same functionality as the original contract \cite{ethereum}. We use this concept of smart contract generation for customized data sharing abilities between different data providers and data requesters.

For scenarios where either the provider or the requester breaches the agreement (i.e., the provider sends corrupted data, the requester sends partial payment, the provider sends no data, etc.), we provide enforcement capabilities by allowing specific pre-designated third party auditors to securely vote on the severity of the breach and decide on the penalty according to the initial data sharing agreement.

Our contributions can be summarized as follows:

\begin{itemize}
\item We provide a novel enforcement framework for data sharing agreements using smart contracts where penalties can be imposed using a carefully designed voting mechanism.
\item The framework is flexible in that it is able to support different kinds of terms and conditions associated with data sharing agreements.
\item The framework supports decentralized data storage where the data providers continue to store and control their own data.
\item Nuanced authentication and authorization can be supported by our encryption scheme.
\item The framework is efficient in that it does not use unnecessary address space on blockchain and contracts are set to self-destruct.
\item We evaluated the proposed framework using the Ethereum Ropsten Network \cite{ropsten} and showed that proposed design is efficient and low cost.
\end{itemize}

The remainder of the paper is structured as follows: Section \ref{related-work} compares and contrasts our work with previous work in data sharing using blockchain technology. Section \ref{overview} provides illustrations and a detailed overview of the system architecture. In section \ref{control-flows}, we provide the steps users take to use our system. In section \ref{breach}, we provide the details of how the data sharing agreement is captured into a voting contract in order to make decisions on agreement breaches. In section \ref{digital-conversion}, we provide the pseudocode of the working contract along with a simple verification example. We discuss experimental results of the system in section \ref{experiments}. In section \ref{summary}, we summarize the advantages of our framework and conclude the paper.

\section{Related Work} \label{related-work}
To our knowledge, much of the past literature in blockchain data sharing has been performed by health care providers, biomedical researchers, and medical schools, so we discuss this field in details below.

In terms of genomic data, Koepsell has attempted to use a multichain blockchain platform as a medium to share anonymized genomic data for scientific research \cite{koepsell}. This blockchain implementation is payment enabled, and only allows researchers to view the metadata of the genomes rather than personal information of the genome donors. However, this blockchain also only allows researchers to sign up as nodes in the private blockchain. Data providers cannot sign up for this blockchain and must instead provide their genomic data to a researcher on the network. Researchers then upload genomic data they have collected to share with other researchers in the network. There is no implementation of constraints on data sharing such as provided by our framework, and every researcher on the blockchain has access to data shared by every other researcher.

In the field of medical imaging data, UCLA has developed their own implementation of a private chain to share data \cite{ucla}. Data providers upload their own data, which is then shared with physicians and personal heath record vendors. Although this implementation does support constraints on the data sharing that limit which vendors can view what data, it requires data providers to write their own data access smart contracts and deploy them on the blockchain. This is a complex task with a steep learning curve, so many data providers might become discouraged and less willing to share their data on this platform.

Medrec is a platform for medical record sharing based on the Ethereum blockchain backbone, and allows health care providers to share medical records between themselves and with their patients, letting patients view an aggregate of their personal medical records from all health care providers they have visited \cite{medrec}. Healthcare providers maintain control over the patient data stored on their servers, but enter “patient-provider relationships” for data access via smart contracts. These relationships are similar to the generalized data sharing agreements our system enables. All patient data access and data sharing by health care providers are logged in an immutable ledger. Medrec also provides the possibility of payment for nodes working to confirm the ledger’s validity, rewarding these nodes with anonymized medical metadata for research purposes. More significantly, Medrec provides a clean web interface for both patients and health care providers to view and share medical records, abstracting away the underlying blockchain and smart contract implementation. This implementation solves the ease-of-use challenge. However, data storage is not decentralized, with all patient data being stored on the health care providers local database. In Medrec's case, the blockchain only serves as a simple access control checker versus an enforcement mechanism such as our framework.

Many of these previous works provide a means to restrict data access based on preferences set by the data provider, but none give the ability to indicate a suspected breach with the agreement after the data has already been shared. This makes these systems not very robust as there is no recourse to data providers or requesters when agreements are breached. 

Although each of the related works discussed above partially solve challenges related to sharing of sensitive data, none of them address all challenges within a single implementation. This project extends the above solutions for blockchain based data sharing and not only provides an easy-to-use contract factory API for generating data sharing contracts on the blockchain, but also allows both data providers and requesters a means to indicate that they suspect a breach and enforce the contract terms. Also, data storage is truly  decentralized, with individuals storing their data in their own locations instead of a central database that contains all the data. This makes the system very robust, scalable, and adaptable to different use cases.

\section{Overview of the framework} \label{overview}
Our goal is to provide an efficient, robust, secure, reliable and a novel way to share data on a blockchain while following nuanced data sharing agreements. The first step in sharing data is for the data provider to store encrypted data on a cloud server. The hash of the data is encrypted with a symmetrical key hosted on the cloud. Hashing allows for efficient storage of the data log. This storage needs to happen before a smart contract is initiated to facilitate robustness of the system. We explored the automatic smart contract generation capability provided by the Ethereum blockchain and exploited its properties to have a minimal code structure. The minimal code structure, also called ‘contract factory’, generates data sharing smart contracts on the Ethereum blockchain “on the fly” when two parties wish to share data based on a legal agreement. The terms of the agreement are decided off the chain. Using our Web interface, the parties provide these terms to our framework. In order to generate a custom smart contract for data sharing, the contract factory requires some Ether from the data requester and considers it as a deposit to start the contract. The deposit is divided according to the following rule:
\begin{equation}
    \text{Deposit} = \text{Payment} + \text{Deposit Money} + \text{Gas Money}. 
\end{equation}

\begin{figure}[h]
    \centering
    \includegraphics[width=0.4\textwidth]{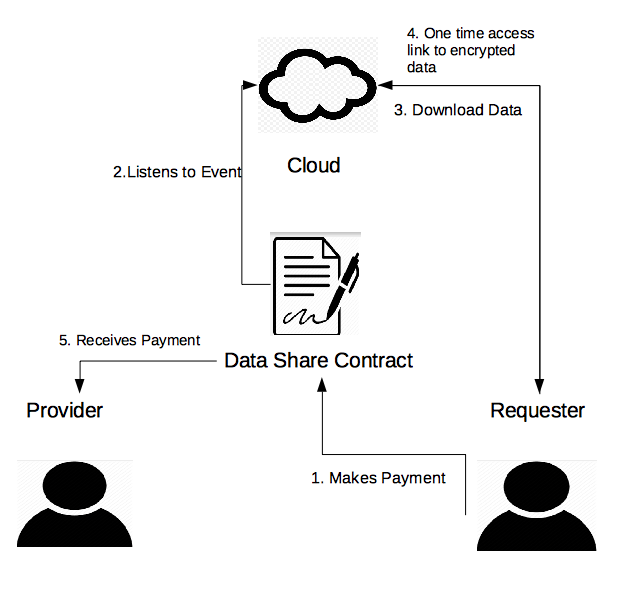}
    \caption{{\bf System Architecture}}
    \label{fig:diagram3}
\end{figure}

The generation of this new smart contract then triggers an event for which the data provider is listening and provides a notification indicating that the provider is required to share the data. When the provider receives this notification, it sends a message including the requester's address/details to the cloud address. The cloud then encrypts the data with the requester’s public key and since the data is already hashed, it gets stored as a one time access link. This link will disappear once it is clicked by the requester. This multi-layer encryption helps in authentication, authorization and secure storage of data over the cloud. The cloud sends the link to the data provider to forward to the requester. Figure \ref{fig:diagram3} illustrates the overall system architecture.

\begin{figure}[h]
    \centering
    \includegraphics[width=0.4\textwidth]{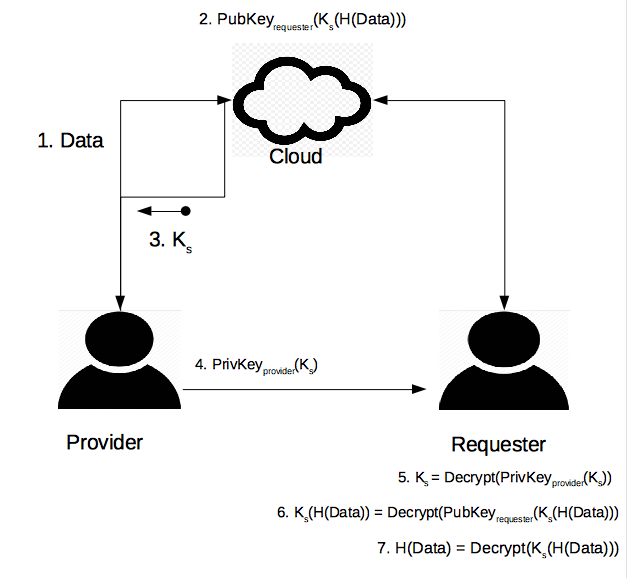}
    \caption{{\bf Role of encryption in our framework:} This multi-layer encryption helps in authentication, authorization and secure storage of data over the cloud. H(Data) is the hash of the data. $K_{s}$ is the symmetric key to encrypt the hash of the data in the cloud. Decrypt is the function used by the requester to use the necessary key he possesses to decrypt the data sent.}
    \label{fig:ecryptiondiagram}
\end{figure}

In order to delegate the data link to the requester, the provider has to pay the smart contract a deposit. In case a breach is to be detected in later stages of the sharing agreement and the provider is found guilty, the smart contract will use the Ether paid by the provider as a deposit to the data requester and the provider will lose its deposit. When the provider sends the link to the requester, it sends the symmetrical key to the data encrypted with the provider’s private key. When the link is sent to the requester over the blockchain, the smart contract dispatches the payment to the data provider without any delay. The requester then clicks the one time link and retrieves the symmetric key using the provider’s public key as a first step. The provider then uses its private key to decrypt the data link as a second step and then uses the symmetric key retrieved from the first step to decrypt the data at the third step. Once the data is retrieved the link will be inaccessible. This completes the data sharing agreement. The encryption process is shown in Fig. \ref{fig:ecryptiondiagram}.

The generated data sharing contract stays in the history of all the three parties including the data provider, the data requester and the cloud for a period specified in the original agreement. The data sharing contract can be destroyed prior to the contract time based self-destruct if both the parties agree on its destruction indicating that the data sharing process has completed successfully and no breach has occurred. Once the data sharing contract is complete, prior to destruction, the smart contract will dispatch the remaining deposit/escrow paid by the participating entities back to them if no breach has occurred or if a breach has occurred and there is escrow remaining with the data sharing contract. \\

\begin{figure}[h]
    \centering
    \includegraphics[width=0.4\textwidth]{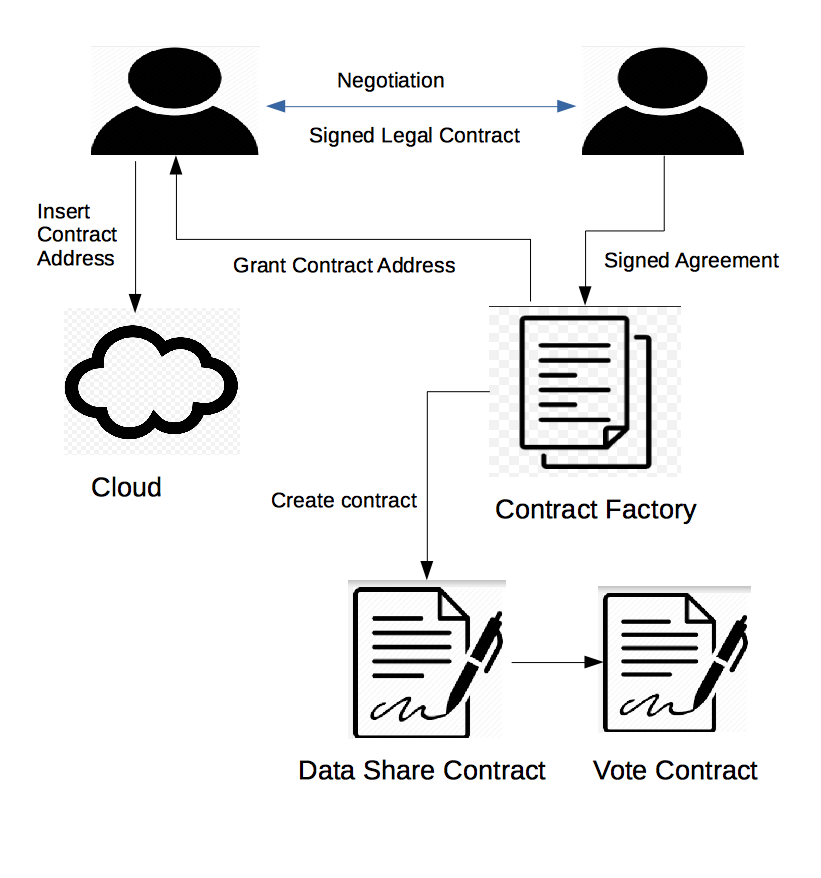}
    \caption{{\bf Control Flow:} The data provider stores the data in the cloud. The data requester and data provider negotiate a data sharing agreement off the chain and use our Web interface to generate a custom smart contract. If either party suspects a breach, they can use the smart contract to spawn a voting contract to investigate the breach and penalize the party causing the breach.}
    \label{fig:diagram1}
\end{figure}

\section{Control Flows} \label{control-flows}

As seen in Fig. \ref{fig:diagram1}, there are several flows in our architecture namely (i) requesting data, (ii) advertising data, and (iii) breach suspected.

\begin{itemize}
\item{Requesting data:} In this case, the data requester will retrieve the data provider’s address from the cloud and send a message to the data provider off the chain to request confidential data. The data requester will send a packet designating the data share contract parameters to the data provider over https, thus making it secure. Once the provider receives the contract parameters, he  decides if he wants to agree to the request or negotiate the contract parameters by modifying the packet received and re-sending it via https. This back-and-forth transition of the “contract parameters” in the form of a packet or also termed as “contract packets” will keep going on until both parties agree on a common set of parameters. Once the agreement is set, the contract is created. The first step is to make the data requester pay for the data where the payment is made to the contract’s address. The data provider receives a notification regarding the payment being made to the contract’s address. The data provider will now send a notification to the cloud to share the data to the provider. The cloud using its encryption policies sends the link of the data to the data requester. Once the requester clicks on the one-time access link, the link will be disabled and depending on the event generated, the payment which was made to the contract’s address by the data requester will be transferred to the data provider’s account. Certain amount of gas will be utilized in the transaction processes, but that amount of gas will be compensated for by the extra amount asked by the contract for the data requester to make a request. 
\footnote{The reason why there is a back-and-forth transition of contract parameters is because we would not want to start the contract with undecided parameters since it takes approximately 30-50 seconds for a contract to be created over the Ethereum network \cite{ethereum-transaction-times}. Also, creating empty contracts which have no potential of being used is a bad practice. It uses up blockchain memory utilizing miners’ time which could have been spent in them mining legitimate contracts and legitimate transactions.}\\

\item{Advertising data:} In this case, when a data provider registers with the system, his address will be sent to the cloud’s address and stored in the cloud database, so that anyone trying to establish a connection with an acting party on the system will go through the cloud to retrieve the address. This way the data provider advertises his address to the set of data requesters, so that they can start sending “contract packets” to the data provider off the chain for an agreement to be established. \\

\item{Suspected breach:} A breach suspected by a data requester can be in multiple forms. Some common forms of a breach would be (a) The data is incomplete. (b) The data is incorrect. (c) The payment extracted is too much. A breach suspected by a data provider can be in multiple forms. Some common forms of such a breach would be (a) The payment is not correct according to the pre-decided terms. (b) The payment is not made at all. (c) The requester violated the policy by sharing the data with a third party. In such cases, the requester or provider will start the voting contract. Depending on the decision on the voting contract, the data share contract will penalize the breacher. In order to do this, as a part of the data share agreement before the data share contract is even created, a fixed deposit will be extracted from both the provider and the requester, in case a breach happens in the future. Once it is confirmed that a breach has occurred, the penalization process will involve sending the ether deposited by the breacher and the victim collectively to the victim. The breacher will not get a refund back. In case a breach never occurs, since the data share contract automatically destroys itself after a certain period of time, the deposit ether will be refunded to each party on contract destruction thus indicating a successful completion of the data share contract. The only drawback of this strategy is that the deposit will remain with the contract while the contract exists in the network or till a breach issue is raised and the breacher is identified. In case of a false accusation, nothing will happen. However, the issuer of the accusation will have to pay an initial fee in terms of gas in order to start the Voting contract to vote on the breach. This can be analogous to the payment of certain legislative fees in order to initiate court proceedings in the real world. 
\end{itemize}

\section{Agreement Breach} \label{breach}
If either party suspects a breach within the lifespan of the data sharing agreement, they can call the data sharing contract and spawn a voting contract to investigate the breach. The breach can occur due to a multiple of reasons. Some of the reasons include not receiving payment, receiving incomplete data, receiving tampered data, data coming from an invalid source, invalid keys, misuse of data, etc. When submitting a voting request, the breach victim writes a description of what they suspect the breach to be, and the compensation that they are seeking for this breach. This payment amount can be specified either when the breach contract is being created, or default to the agreed upon breach penalty in the initial legal agreement. We leave this parameter variable to account for the different severity of possible breaches. 

\begin{figure}[h]
    \centering
    \includegraphics[width=0.5\textwidth]{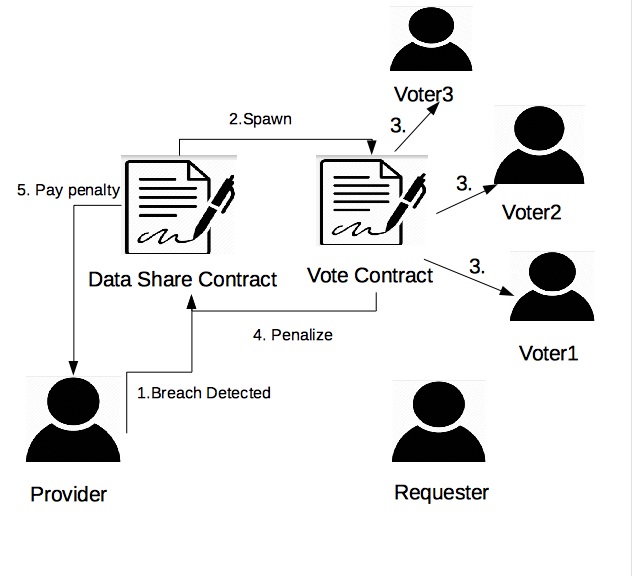}
    \caption{{\bf Agreement breach process:} A voting contract is spawned that leads to penalties if the violation is agreed upon by the voters.}
    \label{fig:breach}
\end{figure}

The CongressFactory smart contract provides all necessary functions for the individual data sharing contracts to call to spawn voting contracts, should a breach ever occur. The CongressFactory takes in as parameters a list of arbiter account addresses, the time voting should be open for, the monetary compensation to be paid to the victim should a breach have indeed occurred, and a description provided by the victim on why they suspect a breach. The CongressFactory will be called by the data sharing contract to spawn a Vote Contract, as shown in Figure \ref{fig:breach}.

After a vote contract is generated and automatically deployed to the blockchain, a request to vote is sent out to all arbiters listed in the vote contract. The arbiters then have until the voting deadline to vote on whether or not a breach has occurred by the checking the blockchain log based on the description of the breach provided by the breach victim. Once the voting deadline has closed, the votes are tallied.

Depending on the decision of the breach, a penalty will then be charged to the violator and the violator’s deposit will be transferred to the victim’s account. The victim will also receive the deposit put into smart contract by itself prior to participating in the smart contract. This completes the Breach Contract/Vote Contract. Once the Breach Contract reveals its decision, it will destroy itself since its purpose is fulfilled and its existence would only be using address space. If another breach is detected within the time period defined by the contract, then another breach contract can be spawned, voting can be redone and the decision can be dispatched again. The voters are decided prior to the generation of the smart contract and off the chain as a part of the original agreement.

\begin{figure}[t!]
    \centering
    \includegraphics[width=0.5\textwidth]{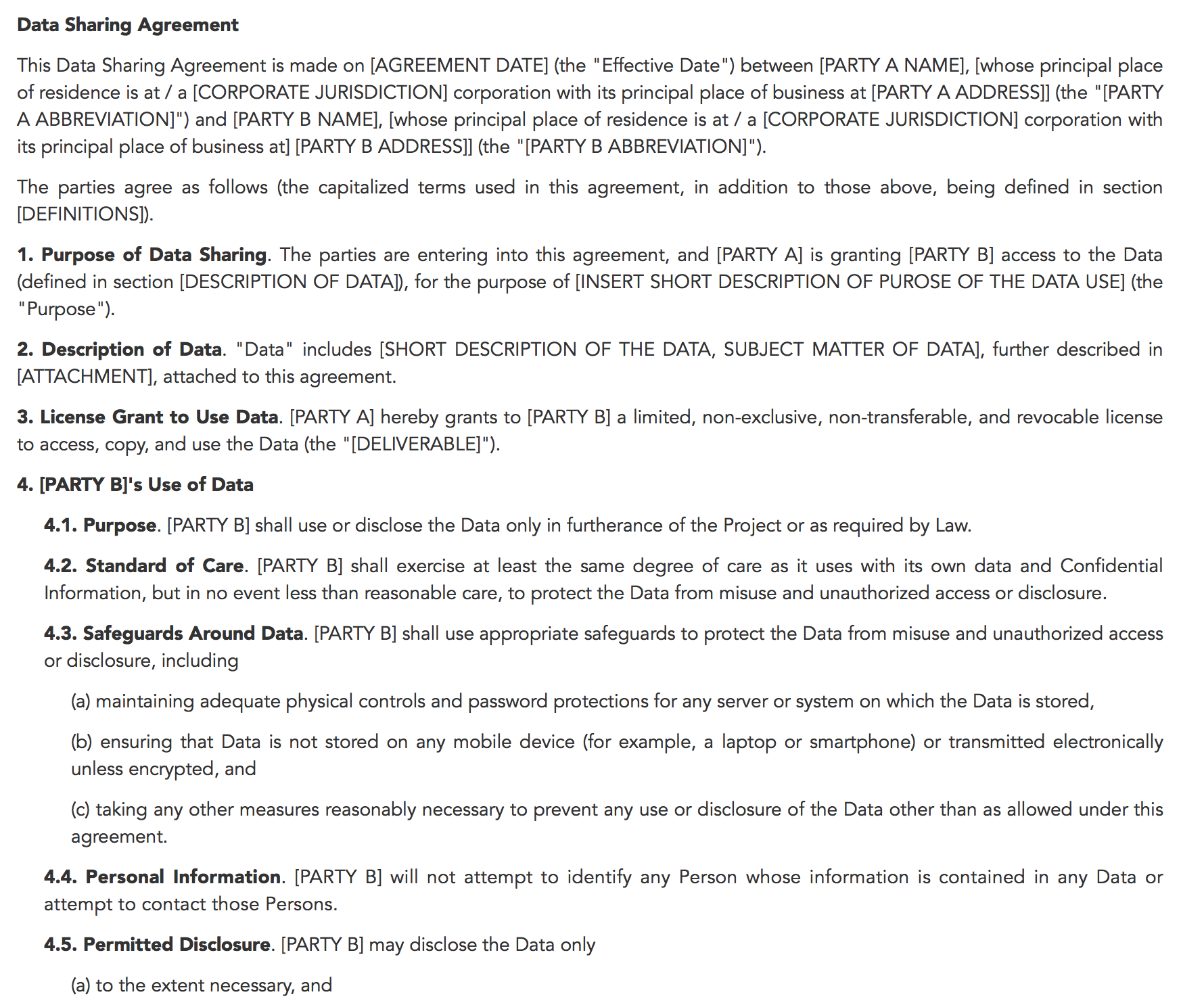}
    \caption{{\bf Legal Data Sharing Agreement:} A short extract of a legal data sharing agreement that includes terms such as purpose of sharing, description of data, and purpose of use.} 
    \label{fig:diagram21}
\end{figure}

\begin{figure*}[t]
    \centering
    \begin{subfigure}[t]{0.35\textwidth}
        \centering
        \includegraphics[height=2in] {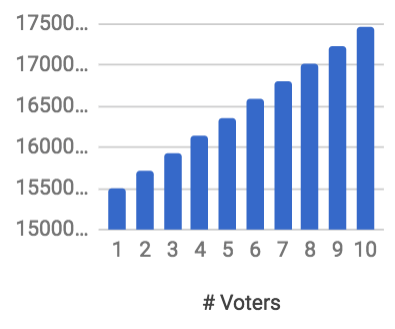}
        \caption{Gas Consumption}
        \label{fig:ds-gasconsumption}
    \end{subfigure}%
    ~ 
    \begin{subfigure}[t]{0.3\textwidth}
        \centering
        \includegraphics[height=2in]{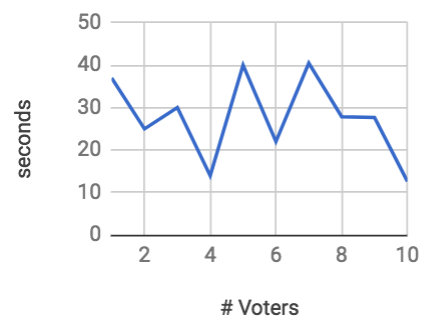}
        \caption{Time Taken for a Specific Instance}
        \label{fig:ds-timetaken}
        \end{subfigure}%
    ~ 
    \begin{subfigure}[t]{0.3\textwidth}
        \centering
        \includegraphics[height=2in] {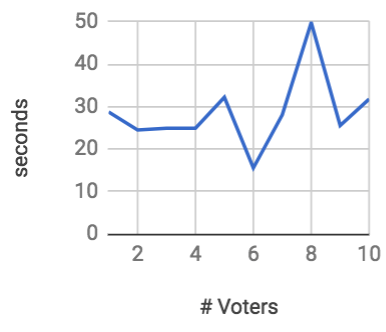}
        \caption{Average Time Taken}
        \label{fig:ds-averagetimetaken}
    \end{subfigure}%
    
    \caption{\emph{Experimental results of Data Sharing Smart Contracts.}}
\end{figure*}

\begin{figure*}[t]
    \centering
    \begin{subfigure}[t]{0.35\textwidth}
        \centering
        \includegraphics[height=2in] {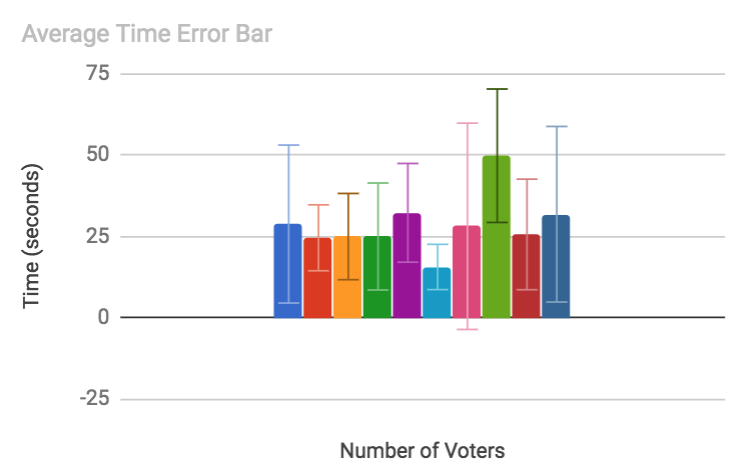}
        \caption{Error Bar}
        \label{fig:ds-errorbar}
    \end{subfigure}%
    ~ 
    \begin{subfigure}[t]{0.3\textwidth}
        \centering
        \includegraphics[height=2in]{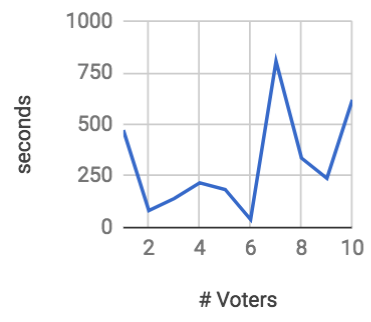}
        \caption{Time Taken Variance}
        \label{fig:ds-timetakenvariance}
        \end{subfigure}%
    ~ 
    \begin{subfigure}[t]{0.3\textwidth}
        \centering
        \includegraphics[height=2in] {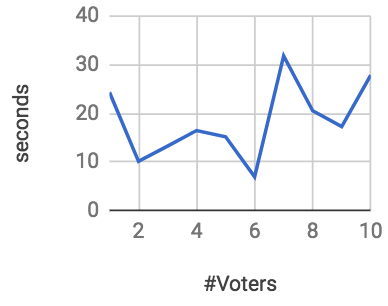}
        \caption{Time Taken Standard Deviation}
        \label{fig:ds-timetakendeviation}
    \end{subfigure}%
    
    \caption{\emph{Experimental results of Data Sharing Smart Contracts.} }
\end{figure*}

\begin{figure*}[t]
    \centering
    \begin{subfigure}[t]{0.35\textwidth}
        \centering
        \includegraphics[height=2in] {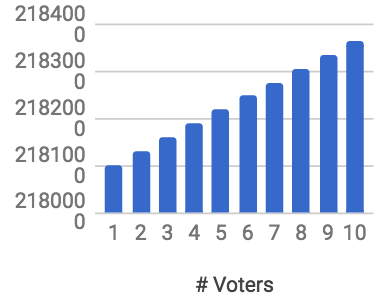}
        \caption{Gas Consumption}
        \label{fig:ds-gasconsumption}
    \end{subfigure}%
    ~ 
    \begin{subfigure}[t]{0.3\textwidth}
        \centering
        \includegraphics[height=2in]{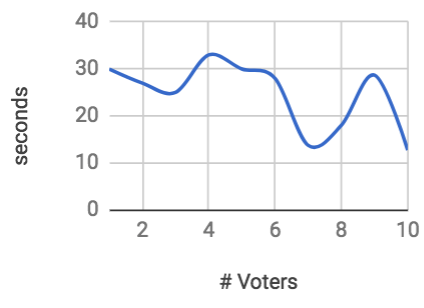}
        \caption{Time Taken for a Specific Instance}
        \label{fig:ds-timetaken}
        \end{subfigure}%
    ~ 
    \begin{subfigure}[t]{0.3\textwidth}
        \centering
        \includegraphics[height=2in] {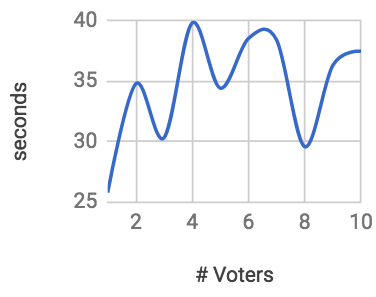}
        \caption{Average Time Taken}
        \label{fig:ds-averagetimetaken}
    \end{subfigure}%
    
    \caption{\emph{Experimental results of The Congress Smart Contracts.}}
\end{figure*}

\begin{figure*}[t]
    \centering
    \begin{subfigure}[t]{0.35\textwidth}
        \centering
        \includegraphics[height=2in] {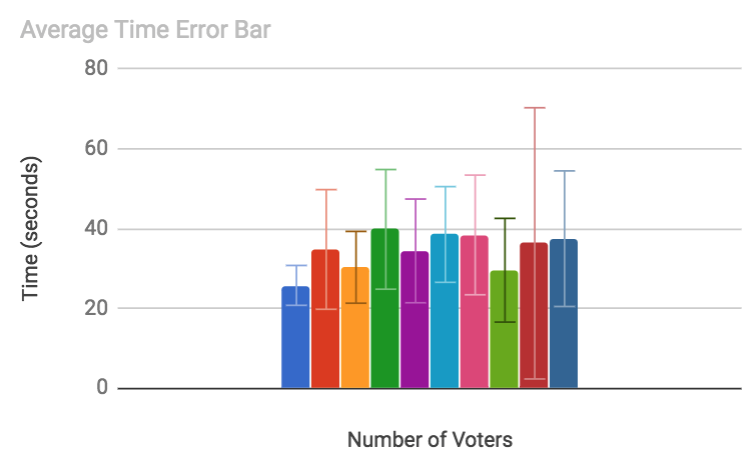}
        \caption{Error Bar}
        \label{fig:ds-errorbar}
    \end{subfigure}%
    ~ 
    \begin{subfigure}[t]{0.3\textwidth}
        \centering
        \includegraphics[height=2in]{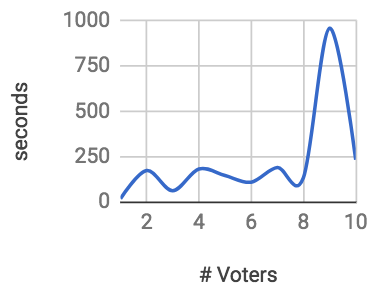}
        \caption{Time Taken Variance}
        \label{fig:ds-timetakenvariance}
        \end{subfigure}%
    ~ 
    \begin{subfigure}[t]{0.3\textwidth}
        \centering
        \includegraphics[height=2in] {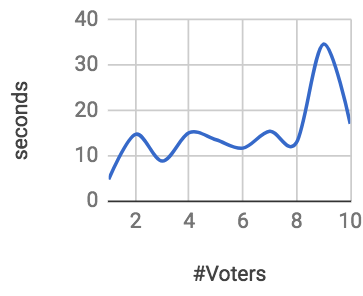}
        \caption{Time Taken Standard Deviation}
        \label{fig:ds-timetakendeviation}
    \end{subfigure}%
    
    \caption{\emph{Experimental results of The Congress Smart Contracts.} }
\end{figure*}

\section{From Data Sharing Agreements to Smart Contracts} \label{digital-conversion}

Data sharing agreements define the conditions under which the sharing may occur (see Fig. \ref{fig:diagram21} for an example) and needs to be performed by lawyers. In many cases, these data sharing agreements are hard to track, manage or enforce and usually require expensive litigation. Our framework is flexible enough to capture and enforce most terms of formal agreements such as purpose of sharing, purpose of use and disclosure. The parameters required for our data sharing smart contracts are data requester's name and address, data provider's name and address, payment, breach condition, voter list, number of votes for quorum, voting time and voting margin. Our Web interface allows the parties to provide these parameters as well as additional terms that are then passed to the contract factory for generating a custom smart contract. If a breach is suspected, these terms are currently manually checked by the specified voters. We are investigating mechanisms to automate this checking of agreement terms as an option to voting.


\section{Implementation Details} \label{implementation}
Ganache-CLI, formerly known as testrpc, was used as a test network for building the initial prototype \cite{testrpc}. We also developed a front end to communicate with the test network. NodeJS \cite{notejs} was used the Javascript run-time environment used to compile and test the functions of the solidity code. The smart contracts were programmed in Solidity \cite{solidity}. The test network provided default gas and ether were used as a simulated money to change the state of the blockchain depending on the function call. Our framework relies on voting for adjudication purposes. Public key cryptography and Symmetric Key cryptography were used for encryption and signing purposes.

The Contract Factory as a whole was deployed on the local network using Ganache-CLI and the public network using Ropsten. Truffle was used as the framework to compile, test, migrate and analyze the deployed contracts \cite{truffle}. A client-server model for the application side of the system was developed following the MVC architecture (Model-View-Control) to make the system more realistic. Technologies like Javascript, NodeJS, ExpressJS and HTML were used along with RESTful API to make http calls to the server from a client. web3.js was used as the sole library to interact with the Ethereum ecosystem. The ContractFactory was deployed along with the CongressFactory as a one-stop shop for the system. The deployment of the contracts as a single entity worked out fine over the local network, however on deployment of the contracts over Ropsten, a gas limit issue was faced. The issue was as follows: On deployment of the contract with a gas limit of anything under 4712388, gives an out of gas error, which means that it uses up the entire gas provided by the contract owner and needs some more. On deployment of the contract with a gas limit of above 4712388, it gives a “Block Gas Limit reached” error which means that Ropsten as a network accepts only transactions whose gas limit is under the specified range. This poses as a deadlock for developers trying to deploy heavy contracts over the Ropsten network. The only solution out is to separate the contract into two small contracts namely, ContractFactory and CongressFactory where the ContractFactory is for sharing data and CongressFactory is for making decisions whether a breach has occurred and enforcing payments from the policy breaker. 


The newly deployed contracts now take the right amount of gas which are as follows: 3047711 for the ContractFactory and 2913993 for the CongressFactory. The addresses of the deployed contracts are as follows: \\
ContractFactory: \\
\texttt{0x7c2842d44e7d4535b50f1c975d5cb04f5324ac8f}\\
CongressFactory: \\
\texttt{0x31e6b4f2be5aec35390f682e9369f385c4daa602}\\


\section{Experiments and Analysis} \label{experiments}
As a part of the analysis, in order to understand the speed of the system, contracts were created in batches assuming if a data requester would want to create multiple contracts with multiple data providers depending on the agreement they have off the chain. 

While performing system analysis, the following parameters were taken into consideration: \\1. Total amount of gas creation.\\ 2. Total time  it takes to deploy the contract. \\3. Variation of the amount of gas it takes in order to deploy the contract. \\4. Variation in amount of gas consumed on changing parameters. \\5. Variation in amount of gas consumed on changing number of voters.





Since the actual time taken from the the deployment of the contract to the time the contract is mined by miners is not determined by the network itself, a measure was taken in the form of number of seconds it takes from deployment to mining. On preliminary analysis, it is obvious that since the time taken from deployment to mining of contracts is really dependent on whether the miners want to mine the transaction and also when the miners would mine the transaction, the time measured should not depend on the contract parameters. It should be arbitrary. However to measure the efficiency of the system, each contract with a specific set of parameters were deployed five times each to calculate the average, and standard deviation of the time taken from deployment to mining. It is found that the average time taken by the data share contract to be mined is approximately 20-50 seconds. Also, the average time taken by the CongressContract is approximately 25-40 seconds. 

Fig 6(a) and Fig 8(a) depict the gas utilized by the network in order to deploy the smart contract based on the number of voters supplied in the DataShare contract parameters. As you can see, the range to deploy a data share contract starts from 1549929 wei which is the gas consumption for having 1 voter for the Congress contract to 1745750 wei which is the gas consumption for having 10 voters for the Congress contract. However, the range to deploy the Congress contract seems to be higher than the corresponding DataShare contract. The range to deploy the corresponding Congress contract starts from 2181014 wei for having 1 voter in the Congress contract to 2183669 wei for having 10 voters in the Congress contract. The slope of each graphs are a constant which means that the gas consumption will increase proportionally with the increase in the number of voters. 

Fig 6(b) and Fig 8(b) shows the amount of time it takes in seconds for an instance of a DataShare contract and Congress contract respectively to be mined by a miner in the Ropsten network since the time of deployment. The graphs are not a straight line which is understandable because of the hypothesis that it depends on the miners’ interest to mine the contracts. They can take arbitrary time to realize completion. 

Fig 6(c) and Fig 8(c) are showing the average time taken in seconds for each set of voters when the same contract is deployed 5 times consecutively. 

Fig 7(a) and Fig 9(a) portrays the error bar of the time taken for each set of voters. The maximum deviation from the average is shown when 7 voters are used to create the DataShare contract and the minimum deviation is shown when 6 voters are used to do the same. On the other hand the maximum deviation from the average is for when 9 voters are used to create the corresponding Congress contract and the minimum deviation is for when just 1 voter is included. 

Fig 7(b) and Fig 7(c) display the variance and standard deviation respectively for a DataShare Contract to be deployed which is bound to be arbitrary since the time taken itself is arbitrary. 

Similarly, Fig 9(b) and Fig 9(c) display the variance and standard deviation respectively for a Congress Contract to be deployed which is also arbitrary since the time taken depends on the network.

\section{Summary and Future Work} \label{summary}

As data becomes increasingly important for businesses, governments, and research, better and more efficient ways need to be found for sharing this data. Although data sharing is required, in order to meet privacy, security and regulation requirements, data providers need to specify nuanced conditions under which the sharing occurs. Both the specification and enforcement process of these data sharing agreements is manual and done by lawyers and through litigation. We propose an automatic way for tracking, managing and especially enforcing such data sharing agreements using smart contracts and blockchain technology. Our framework generates smart contracts from parameters similar to those found in legal data sharing agreements. The terms in this agreements are automatically enforced by the system and monetary punishment can be employed using secure voting. 

We are interested in exploring several future steps. In order to increase trust in our framework, we need to be able to verify the smart contracts and prove that they do what the data sharing parties expect. We are also investigating easy to understand languages and schemas that can be compiled down to Solidity smart contracts. We would like to modify our Web front end to accept digitally signed agreements written in these languages, extract the relevant terms and generate the appropriate smart contracts. This would not only make verifiability easier but would make for an end-to-end automatic enforcement framework. 

\bibliographystyle{ieeetr}
\bibliography{references}



\end{document}